\documentclass{article}[12pt]

\begin{document}


%
%

\title{Ultralocal Energy Density in Massive Gravity}

\author{Vladimir O. Soloviev
and Margarita V. Tchichikina}

\maketitle

\begin{abstract}
We provide a space-time covariant Hamiltonian treatment for a finite-range gravitational theory.  The  Kucha\u{r} approach is used to demonstrate the bimetric picture of space-time in its most transparent form. 
This Hamiltonian formalism 
is applied for the straightforward realization of the Poincar\'e algebra in Dirac brackets. It uncovers the simplest form of the Poincar\'e generators expressed as spatial integrals of  ultralocal quantities constructed pure algebraically by means of the two space-time metrics. 

\end{abstract}


\section{Introduction}

For many reasons, both observational and motivated by pure theory,
modifications of the  gravitational theory attract a lot of attention now. In
particular, we mean those changes in the standard model of
gravity  that modify its infrared behavior. It is possible to achieve the transformation of the massless theory into the massive one 
in different ways. The discussion may be found, for example, in articles \cite{Rubak,DDGV,DK,Bek,Blas,MMMM}. In this  paper we deal with a variant of gravitational theory which is historically connected with works~\cite{Rosen,FMS,LogMe}.

On the ground of the fact that the linearized kinetic energy contains a negative mode~\cite{BD} sometimes it is said that this model does not deserve futher study. But as some other authors~\cite{Visser,BG,Petrov,PS} we can not agree with this statement.  It is necessary to study the behaviour of solutions not only in linear case, but in the full theory, as it can be rather different.

 The content of the theory becomes in many aspects more transparent after expression in the  general language  of Hamiltonian formalism. With this purpose we demonstrate here the construction and elementary applications of the canonical formulation for the Relativistic Theory of Gravitation (RTG)~\cite{Log}. We stress that our main interest is in the generally covariant canonical formalism developed by Kucha\u{r}~\cite{Kuchar} and not in the ADM variant~\cite{ADM}. An essential difference of Kucha\u{r}'s approach is that four components of (pseudo)Riemannian metric tensor and four lapse and shift functions are not identified and play different roles. Precisely, the momenta conjugate to the four metric components are zero and that leads to four primary constraints, whereas those metric components are persent in the secondary constraints. All the four pairs of variables are excluded after solving the second class constraints, whereas lapse and shift functions stay in the Hamiltonian, providing its general coordinate invariance and freedom to choose the foliation of spacetime by spatial hypersurfaces.  Surely both the Kucha\u{r} approach and ADM one are based on the fundamental Dirac results, presented in brilliant form in his Yeshiva Lectures~\cite{Dirac}.  Here we apply, in fact, just in the same order as in the book~\cite{Dirac}, both the variant called the quantization on curved surfaces and the other one called the quantization on flat surfaces.

The bimetric view of General Relativity (GR) first appeared in publication~\cite{Rosen}. After expressing a bimetric theory in the Hamiltonian form we immediately encounter with  the causality issue. In the gravitational theory~\cite{Log} discussed here this problem is solved by introducing an extra  causality postulate.  This postulate requires that the null cone defined by the dynamical (pseudo)Riemannian metric should be inside or coincide with the null cone defined by the flat background metric. Then any hypersurface which is spacelike according to the flat metric is also spacelike in the (pseudo)Riemannian metric.  As covariant Hamiltonian approach allows a free choice of time one always can take it so that its direction will be timelike in both metrics. We demonstrate that the total energy density is proportional to the square of the unit normal to the constant time hypersurface. Of course, the unit normal is defined on the base of one metric and it is squared with another one. So the time problem and energy problem become united.

At last, most of physicists believe that the gravitational field as other fundamental fields subjects to quantization. The difficulties of metric tensor quantization in General Relativity are well known. Perspectives of quantization of the alternative gravitation theories deserve further study. Canonical quantization is the first in the list of methods developed to unite quantum and classical theory, both historically and practically, so we start here with a discussion of the Hamiltonian formalism also having in mind a program of quantization. In Ref.~\cite{Sol86} the bimetric gravity has been treated  as a zero graviton mass theory, but later that version has been declined. The  new statement of this problem, as will be seen below, leads to rather different results. There are no first class constraints in the theory now, the number of degrees of freedom increases and Poincar\'e invariance  results in ten non-trivial integrals of motion. Some preliminary presentation of this work appeared in Refs.~\cite{SoloChi,SoloChi2}

In Section 2 we introduce our notations and the Lagrangian for the considered theory of gravitation. Then we apply to all tensors and vectors the spacetime covariant $3+1$-decomposition  developed by Kucha\u{r}~\cite{Kuchar}. Spacetime is treated as foliated by one-parametric family of spacelike hypersurfaces. The time here is not a coordinate, but a  parameter continuously numbering hypersurfaces. This formalism exploits simultaneously two coordinate systems: the first is general and arbitrary, the second is induced by the foliation. General spacetime coordinate invariance so is respected. The pecularity of bimetric picture is a strong reqirement that hypersurfaces are to be spacelike in both metrics. There are also two bases for decomposition of tensors corresponding to the two spacetime metrics. These decompositions are applied both to proper spacetime tensors and to tensors depending on the foliation. Practically, more attention is paid to the decomposition based on flat background metric, though sometimes it is rather suitable to use another basis.

In Section 3 $(3+1)$-decomposed tensors are substituted into the Lagrangian and after this transformation of variables the Hamiltonian formalism is constructed. The Lagrangian occurs to be a sum of  two terms, one of them coinsides with the term of General Relativity, the other is constructed from the two metric by purely algebraic means. So, the transformation of variables is not difficult. In contrast to one-metric theories, such as General Relativity, all ten components of (pseudo)Riemannian metric enter the Lagrangian and should be treated as dynamical variables. But velocities of four of them do not enter the Lagrangian and so we get four primary constraints. Conservation of them leads to four secondary constraints and all eight constraints occur second class. It is easy to solve four secondary constraints and express four  variables as functions of other variables. After construction of Dirac brackets all four pairs of above mentioned variables and all constraints may be excluded. In the same time Dirac brackets for the rest gravitational variables coincide with their Poisson brackets in General Relativity. The Hamiltonian of the gravitational field has a standard for field theories on fixed background form, i.e. it is linear in lapse and shift functions.

In Section 4 we take the scalar field as a representative of matter. The $3+1$-decomposition again leads us to the field Hamiltonian linear in lapse and shift functions. The gravitational interaction(minimal interaction) of scalar field is provided by (pseudo)Riemannian metric. Only uniting both Lagrangians we obtain the consistent interaction model. We present the full system of canonical equations generated by the Hamiltonian and Dirac brackets.   

Section 5 is devoted to the Poincare group and to generation of its transformations by the Hamiltonian. As Hamiltonian depends on four functions $N$, $N^i$ which are not dynamical variables the evolution depends on their choice. Starting the evolution from a given hypersurface we  are free to choose the following hypersurfaces (of course by preserving smoothness). The evolution parameter is not always possible to be treated as some kind of ``time'', in general, it is simply a parameter of some smooth coordinate transformation. In particular, the Hamiltonian can generate purely spatial coordinate transformations in a hypersurface if we put lapse function to zero. We start with a calculation of Dirac braket for two Hamiltonians with different lapse and shift functions. This bracket almost gives the celebrated ``hypersurface deformation algebra'' derived by Dirac. For the full correspondence the Hamiltonians should not contain other functions which are not canonical variables then lapse and shift. In our case such a function is a spatial  metric induced on  hypersurfaces by the flat background metric of spacetime. Theories with the background metric may be extended by introduction of four pairs of canonical variables: embedding variables and their conjugate momenta, then the exact Dirac formulas are valid. Also these formulas are valid in General Relativity because it does not contain any background metric. In our case the algebra appears when we restrict ourselves by transformations which are motions for the background metric (Killing vectors). For the flat background they are transformations from the Poincar\'e group. In order to preserve not only spacetime background metric but also spatial metric induced by it on hypersurfaces it is necessary to restrict formalism by those which are flat in background metric. And the last restriction is not necessary --  for simplicity we allow only Cartesian coordinates on the hypersurfaces.

If we put a requirement that the empty flat space has exactly zero energy density, then the Poincar\'e algebra will have a central extension and central charges. In other case we will have an algebra without central charges but the empty Minkowski spacetime will have  some small constant positive energy density. 

We should like to draw attention to the fact that energy and momentum densities are ultralocal here and this gives  a straightforward connection between the sign of total energy density and the bimetric causal structure of spacetime: energy density occurs proportional to the square of unit normal to the hypersurface. Of course, this unit normal should be defined by means of background flat metric and its square is calculated by means of another, (pseudo)Riemannian spacetime metric. The reasonable choice of foliation, therefore, should provide us with positivity of the total energy density for gravitational field and matter.

\section{The RTG Lagrangian, 3+1-decomposition of tensors and a new form of the Lagrangian density}
We start with the RTG Lagrangian~\cite{Log},
which allows in more or less standard way to derive the Hamiltonian and the Poisson brackets.
The RTG Lagrangian contains nondynamical flat metric 
$h_{\mu\nu}$ together with dynamical pseudo-Riemannian
metric $g_{\mu\nu}$, this is the main difference from the GRT. Taking the velocity of light equal to unity, we can take the Lagrangian density of the gravitational field following Ref.~\cite{Log} as
\begin{equation}
{\cal L}=\frac{1}{16\pi G}\sqrt{-g}R -\frac{m^2}{16\pi G}\left(\frac{1}{2}
h_{\mu\nu}\tilde g^{\mu\nu}-\sqrt{-g}-\sqrt{-h}\right) +\dots, \label{eq:Lagr}
\end{equation}
where dots denote surface terms (4-divergences),
Greek indices take values from 0 to 3, $G$ is the gravitational constant, $g=\det(g_{\mu\nu})$, $\tilde g^{\mu\nu}=\sqrt{-g}g^{\mu\nu}$,
$h=\det(h_{\mu\nu})$, $R$ is the space-time scalar curvature given by metric $g_{\mu\nu}$, $m$
 is the graviton mass. The signature of space-time is $(-1,1,1,1)$.

We can also present Lagrangian density (\ref{eq:Lagr}) in an equivalent (up to divergences) form 
\begin{equation}
{\cal L}=\frac{1}{16\pi G}
\tilde g^{\mu\nu}\left(\Delta\Gamma^{\lambda}_{\mu\sigma}\Delta\Gamma^{\sigma}
_{\nu\lambda}-\Delta\Gamma^{\lambda}_{\mu\nu}\Delta\Gamma^{\sigma}
_{\lambda\sigma}\right)
-\frac{m^2}{16\pi G}\left(\frac{1}{2}
h_{\mu\nu}\tilde g^{\mu\nu}-\sqrt{-g}-\sqrt{-h}\right) +\dots, \label{eq:Lagr2}
\end{equation}
where
\begin{equation}
\Delta\Gamma^{\lambda}_{\mu\nu}\equiv \bar\Gamma^{\lambda}_{\mu\nu}-
\Gamma^{\lambda}_{\mu\nu}
=\frac{1}{2}g^{\lambda\sigma}\left(
D_{\mu}g_{\sigma\nu}+D_{\nu}g_{\sigma\mu}-D_{\sigma}g_{\mu\nu}
\right),
\end{equation}
$\bar\Gamma^{\lambda}_{\mu\nu}$ are the Christoffel symbols of the pseudo-Riemannian metric,
$\Gamma^{\lambda}_{\mu\nu}$ are  Christoffels of the flat metric, and
 $D_{\mu}$ is the covariant derivative compatible with the flat metric.

As independent variables in the action constructed on the base of Lagrangian densities (\ref{eq:Lagr}) or (\ref{eq:Lagr2}), which should be varied, one can take, for example,
 10 components of pseudo-Riemannian metric $g_{\mu\nu}$.
To simplify the formal calculations the following tensor objects will be useful
\begin{equation}
f^{\mu\nu}\equiv\frac{\sqrt{-g}}{\sqrt{-h}}g^{\mu\nu}.
\end{equation}


In constructing the Hamiltonian formalism it is necessary to chose the arrow of evolution, i.e. the physical time  direction. This does not mean violation of the general covariance and  fixing one of space-time coordinates $X^{\alpha}$ as time.  It was firstly shown in Kucha\u{r}'s works~\cite{Kuchar} how to make the canonical formalism space-time covariant. We will follow this approach here, and we will also use some results derived in paper~\cite{Sol88}.

The fixed instant of physical time corresponds to some spacelike hypersurface
\begin{equation}
X^{\alpha}=e^{\alpha}(x^i),
\end{equation}
where $x^i$ are independent coordinates on the hypersurface, Latin indices take values
from 1 to 3. Unlike the case treated in publications~\cite{Kuchar}, we  deal here with two metrics for the spacetime, not one, and we demand that our hypersurfaces are to be spacelike in both metrics.~\footnote{This requirement is justified by the postulate of causality in the RTG.} It means imposing two conditions valid at any point of the hypersurface
\begin{equation}
\gamma_{ij}(x^k)dx^idx^j>0,\quad \eta_{ij}(x^k)dx^idx^j>0,
\end{equation}
where two different induced metrics are involved:
\begin{equation}
\gamma_{ij}=g_{\mu\nu}e^{\mu}_{,i}e^{\nu}_{,j},\quad
\eta_{ij}=h_{\mu\nu}e^{\mu}_{,i}e^{\nu}_{,j}.
\end{equation}
It is evident that in general case metric $\eta_{ij}$, in contrast to metric $h_{\mu\nu}$,
is not flat.

We will assume that the spacetime may be filled with such slices of constant physical time, i.e. presented as one-parametrical family of spacelike hypersurfaces:
\begin{equation}
X^{\alpha}=e^{\alpha}(x^i,t),
\end{equation}
and we can introduce some vector field
\begin{equation}
N^{\alpha}=\frac{\partial e^{\alpha}}{\partial t},
\end{equation}
which is timelike in both metrics of spacetime
\begin{equation}
g_{\alpha\beta}N^{\alpha}N^{\beta}<0,\quad
h_{\alpha\beta}N^{\alpha}N^{\beta}<0.
\end{equation}

In order to provide $3+1$-decompositions of spacetime tensors we are to have a basis related to a fixed time hypersurface. For example, we may use for this purpose four spacetime vectors $(N^{\alpha},e^{\alpha}_{,i})$. But really we need two bases: one connected to metric $g_{\mu\nu}$, and another connected to metric $h_{\mu\nu}$. This doubling occurs when we go to lower indices
\begin{equation}
N_{\beta}=h_{\beta\alpha}N^{\alpha},\quad
\bar N_{\beta}=g_{\beta\alpha}N^{\alpha},
\end{equation}
and analogously, $e_{\alpha i}$ and $\bar e_{\alpha i}$ appear.
It is more suitable from the technical view to introduce two other bases, where  unit normals to the hypersurface are taken as timelike vectors
 $(n^{\alpha},e^{\alpha}_{,i})$ and
$(\bar n^{\alpha},e^{\alpha}_{,i})$. They are defined by the following conditions
\begin{equation}
h_{\alpha\beta}n^{\alpha}e^{\beta}_{,i}=0,\quad
h_{\alpha\beta}n^{\alpha}n^{\beta}=-1,
\end{equation}
\begin{equation}
g_{\alpha\beta}\bar n^{\alpha}e^{\beta}_{,i}=0,\quad
g_{\alpha\beta}\bar n^{\alpha}\bar n^{\beta}=-1.
\end{equation}

Now we are able to apply 3+1-decomposition to different tensors, for example,
\begin{eqnarray}
N^{\alpha}&=&Nn^{\alpha}+N^ie^{\alpha}_{,i}=\bar N\bar n^{\alpha}+\bar N^i
e^{\alpha}_{,i},\nonumber\\
g^{\mu\nu}&=&g^{\perp\perp}n^{\mu}n^{\nu}+g^{\perp j}n^{\mu}e^{\nu}_{,j}+
g^{i\perp}e^{\mu}_{,i}n^{\nu}
+g^{ij}e^{\mu}_{,i}e^{\nu}_{,j}=
(-1)\bar n^{\mu}\bar n^{\nu}
+\gamma^{ij}e^{\mu}_{,i}e^{\nu}_{,j},\nonumber\\
f^{\mu\nu}&=&f^{\perp\perp}n^{\mu}n^{\nu}+f^{\perp j}n^{\mu}e^{\nu}_{,j}+
f^{i\perp}e^{\mu}_{,i}n^{\nu}
+f^{ij}e^{\mu}_{,i}e^{\nu}_{,j}
,\label{eq:3+1}
\end{eqnarray}
where
\begin{equation}
N=-n_{\mu}N^{\mu},\quad N^i=e_{\mu}^{i}N^{\mu},\quad \bar N=
-\bar n_{\mu}N^{\mu},
\quad \bar N^i=\bar e_{\mu}^iN^{\mu},\label{eq:parameters}
\end{equation}
\begin{equation}
g^{\perp\perp}=n_{\mu}n_{\nu}g^{\mu\nu},\quad g^{\perp j}=
g^{j\perp}=-n_{\mu}e_{\nu}^j
g^{\mu\nu},
\quad g^{ij}=e_{\mu}^ie_{\nu}^jg^{\mu\nu},\dots
\end{equation}
It is easy to find the linear relations between the two bases
\begin{equation}
\bar n^{\alpha}=\sqrt{-g^{\perp\perp}}n^{\alpha}-\frac{g^{i\perp}}{\sqrt{
-g^{\perp\perp}}}e^{\alpha}_i,
\end{equation}
and correspondingly the linear relations between the components as
\begin{equation}
\bar N=-\frac{1}{f^{\perp\perp}}\sqrt{\frac{\gamma}{\eta}}N,\quad
\bar N^i=N^i-\frac{f^{\perp i}}{f^{\perp\perp}}N.\label{eq:NN}
\end{equation}

\section{Construction of the canonical formalism for the gravitational field}

In order to transform the Lagrangian density to the desired form (\ref{eq:Lagr})
it is necessary to express in a new way the two terms, where the first one does not contain neither the graviton mass, nor the flat metric. Therefore for the first term it is enough to apply the standard transformations ~\cite{Kuchar} used in GRT. Then up to surface terms we obtain
\begin{equation}
-\frac{N}{16\pi G}
\frac{\gamma}{f^{\perp\perp}\sqrt{\eta}}(\tilde R-{\bar K}^2+{\rm Sp}{\bar K}
^2),
\end{equation}
where $\tilde R$
 is the scalar curvature of the hypersurface, derived by means of metric 
$\gamma_{ij}$,   $\bar K_{ij}$
is the second fundamental form of the hypersurface in pseudo-Riemannian geometry defined by metric  $g_{\mu\nu}$,  ${\rm Sp}{\bar K}^2=
\bar K_{ij}\bar K_{kl}\gamma^{ik}\gamma^{jl}$.
Taking into  account formula
\begin{equation}
f^{ij}=\frac{1}{f^{\perp\perp}}\left(
f^{\perp i}f^{\perp j}-\frac{\gamma\gamma^{ij}}{\eta}
\right),
\end{equation}
and after substitution of decompositions
(\ref{eq:3+1}),          the second term takes the following form
\begin{equation}
-N\sqrt{\eta}\frac{m^2}{16\pi G}\left[-1-\frac{f^{\perp\perp}}{2}+
\frac{f^{\perp i}f^{\perp j}\eta_{ij}}{2f^{\perp\perp}}-
\frac{1}{f^{\perp\perp}}\frac{\gamma}{\eta}\left(\frac{1}{2}
\eta_{ij}\gamma^{ij}-1
\right)
\right].
\end{equation}
It is seen that this expression does not contain any velocities and so it is irrelevant for definition of momenta.

The surface terms in Lagrangian density (total derivatives and spatial divergences) do not influence the symplectic structure and so the Poisson brackets. Boundary conditions at the spatial infinity in their turn are taken so, that  the pseudo-Riemannian metric tends to flat Minkowski metric, and the hypersurfaces tend to hyperplanes. So we arrive at the following form of the gravitational field action
\begin{eqnarray}
S&=&
\int\limits^{t_2}_{t_1}dt\int\limits_{R^3}d^3x
\left(
-\frac{N}{16\pi G}\frac{\gamma}{f^{\perp\perp}\sqrt{\eta}}
(\tilde R-{\bar K}^2+{\rm Sp}{\bar K}^2)
\right.
\nonumber\\
&-&
\left.
N\sqrt{\eta}\frac{m^2}{16\pi G}
\left[-1-\frac{f^{\perp\perp}}{2}+
\frac{f^{\perp i}f^{\perp j}\eta_{ij}}{2f^{\perp\perp}}-
\frac{1}{f^{\perp\perp}}\frac{\gamma}{\eta}
\left(\frac{1}{2}\eta_{ij}\gamma^{ij}-1\right)
\right]
\right).\label{eq:action}
\end{eqnarray}
Here $\gamma_{ij}(x^k,t)$, $f^{\perp\perp}(x^k,t)$, $f^{\perp i}(x^k,t)$ are the independent variables to be varied.

Flat spacetime metric $h_{\mu\nu}(X^{\alpha})$ and functions $e^{\alpha}(x^i,t)$ which parametrize the spacelike hypersurfaces are treated as given and so not to be varied.  Base vectors $n^{\alpha}(x^i,t),
e^{\alpha}_i(x^i,t)$ and vector $N^{\alpha}(x^i,t)$ are in their turn expressed through those functions. Quantities
$\bar K_{ij}(x^i,t)$ are given by the formulas well-known from the canonical formalism of General Relativity
\begin{equation}
\bar K_{ij}=\frac{1}{2\bar N}\left(\bar N_{i|j}+\bar N_{j|i}-\gamma_{ij,0},
\right)
\end{equation}
where functions $\bar N$, $\bar N^i$  in their turn are expressed through $N$, $N^i$
and $f^{\perp\perp}$, $f^{\perp i}$ by equations (\ref{eq:NN}).
Here the vertical bar denotes the covariant derivative determined in the Riemannian geometry of 3-dimensional space by metric $\gamma_{ij}$.
In order to avoid confusion of mixing momenta with the famous mathematical constant $\pi$ let us hide this constant in a new constant $\kappa=16\pi G$.

Starting from action (\ref{eq:action})
we obtain for the conjugate momenta the following expressions:
\begin{eqnarray}
\pi_{\perp}&=&\frac{\partial{\cal L}}{\partial f^{\perp\perp}_{,0}}=0,
\label{eq:constraint1}\\
\pi_{i}&=&\frac{\partial{\cal L}}{\partial f^{\perp i}_{,0}}=0,
\label{eq:constraint2}\\
\pi_{ij}&=&\frac{\partial{\cal L}}{\partial\gamma_{ij,0}}=
\frac{\partial{\cal L}}{\partial\bar K_{ij}}\frac{\partial\bar K_{ij}}
{\partial\gamma_{ij,0}}=-\frac{\sqrt{\gamma}}{\kappa}
(\bar K^{ij}-\gamma^{ij}
\bar K).
\end{eqnarray}
It is evident from the above that equations (\ref{eq:constraint1}) and  (\ref{eq:constraint2})
are primary constraints in Dirac's terminology~\cite{Dirac}, and so these equations are to be added with arbitrary multipliers to the Hamiltonian.
Therefore we obtain
\begin{equation}
{\rm H}=\int\limits_{R^3}d^3x\left(
\pi^{ij}\gamma_{ij,0}-{\cal L} +\lambda^{\perp}\pi_{\perp}+\lambda^i\pi_i
\right),
\end{equation}
where velocities should be expressed through moments as follows
\begin{equation}
\gamma_{ij,0}=\bar N_{i|j}+\bar N_{j_i}+\frac{2\kappa\bar N}{\sqrt{\gamma}}
(\pi_{ij}-\gamma_{ij}\frac{\pi}{2}).
\end{equation}
Now, after finishing this procedure, the Hamiltonian takes the following form (up to surface terms)
\begin{equation}
{\rm H}=\int\limits_{R^3}d^3x\left(
N{\cal H}+N^i{\cal H}_i +\lambda^{\perp}\pi_{\perp}+\lambda^i\pi_i
\right),\label{eq:Ham1}
\end{equation}
where
\begin{eqnarray}
{\cal H}&=&-\frac{1}{f^{\perp\perp}}\sqrt{\frac{\gamma}{\eta}}{\bar{\cal H}}-
\frac{f^{\perp i}}{f^{\perp\perp}}{\bar{\cal H}}_i\nonumber\\
&+&\frac{m^2\sqrt{\eta}}{\kappa}\left[
-1-\frac{f^{\perp\perp}}{2}+
\frac{f^{\perp i}f^{\perp j}\eta_{ij}}{2f^{\perp\perp}}-
\frac{1}{f^{\perp\perp}}\frac{\gamma}{\eta}\left(\frac{1}{2}
\eta_{ij}\gamma^{ij}-1
\right)
\right],\\
{\cal H}_i&=&\bar{\cal H}_i=-2\pi_{i|j}^j,\\
\bar{\cal H}&=&-\frac{1}{\sqrt{\gamma}}\left(
\frac{1}{\kappa}\gamma\tilde R+\kappa(\frac{\pi^2}{2}-\mathrm{Sp}\pi^2)
\right).
\end{eqnarray}

The canonical Poisson brackets given as follows
\begin{equation}
\{F,G\}=\int\limits_{R^3}d^3x\left[\frac{\delta F}{\delta\gamma_{ij}}
\frac{\delta G}{\delta\pi^{ij}}+
\frac{\delta F}{\delta f^{\perp\perp}}
\frac{\delta G}{\delta\pi_{\perp}}
+
\frac{\delta F}{\delta f^{\perp i}}
\frac{\delta G}{\delta\pi_{i}}-(F\leftrightarrow G)
\right]\label{eq:PB}
\end{equation}
let us to present the hamiltonian equations in a standard form:
\begin{equation}
\gamma_{ij,0}=\{\gamma_{ij}, {\rm H}\},\quad
\pi^{ij}_{,0}=\{\pi^{ij}, {\rm H}\},\label{eq:he1}
\end{equation}
\begin{equation}
f^{\perp\perp}_{,0}=\{f^{\perp\perp},{\rm H}\},\quad
\pi_{\perp,0}=\{\pi_{\perp}, {\rm H}\},\label{eq:he2}
\end{equation}
\begin{equation}
f^{\perp i}_{,0}=\{f^{\perp i},{\rm H}\},\quad
\pi_{i,0}=\{\pi_{i}, {\rm H}\}.\label{eq:he3}
\end{equation}

Next we are to check whether the primary constraints (\ref{eq:constraint1}),
(\ref{eq:constraint2}) are compatible with the equations of motion, in order to guarantee that time derivatives $\pi_{\perp,0}$ and $\pi_{i,0}$ are also zero.

As  conjugate variables $f^{\perp\perp}$ and $f^{\perp i}$ enter the Hamiltonian algebraically we obtain secondary constraints in the form of algebraic equations:
\begin{equation}
\frac{\partial{\cal H}}{\partial f^{\perp\perp}}=0,\quad
\frac{\partial{\cal H}}{\partial f^{\perp
i}}=0,\label{eq:secondconstraints}
\end{equation}
which can be solved in elementary way and they give relations:
\begin{eqnarray}
f^{\perp i}&=&\frac{\kappa}{m^2\sqrt{\eta}}\eta^{ij}\bar{\cal H}_j,
\label{eq:f_i}\\
f^{\perp\perp}&=&-\frac{\kappa}{m^2\sqrt{\eta}}\sqrt{
\eta^{ij}\bar{\cal H}_i\bar{\cal H}_j+2\frac{m^2\sqrt{\eta}}{\kappa}
\left[
\sqrt{\frac{\gamma}{\eta}}\bar{\cal H}+\frac{m^2\sqrt{\eta}}{\kappa}
\frac{\gamma}{\eta}\left(\frac{1}{2}\eta_{ij}\gamma^{ij}-1\right)
\right]\label{eq:f_perp}
}.
\end{eqnarray}
It is easily seen from the solved form of the secondary constraints that their Poisson brackets with the primary constraints are nonzero, i.e. all the constraints are second class and so may be excluded by introducing the Dirac brackets. Here Dirac brackets may be obtained from Poisson brackets
 (\ref{eq:PB})
by simple exclusion of terms with variables $(f^{\perp\perp},\pi_{\perp})$  and
$(f^{\perp i},\pi_i)$
\begin{equation}
\{F,G\}_D=\int\limits_{R^3}d^3x\left[\frac{\delta F}{\delta\gamma_{ij}}
\frac{\delta G}{\delta\pi^{ij}}
-\frac{\delta F}{\delta\pi^{ij}}\frac{\delta G}{\delta\gamma_{ij}}
\right].\label{eq:DB}
\end{equation}
After substituting the solutions of constraint equations into the hamiltonian we obtain
\begin{eqnarray}
{\rm H}&=&\int\limits_{R^3}d^3x
\left[
N\left(
\sqrt{
\eta^{ij}\bar{\cal H}_i\bar{\cal H}_j+2\frac{m^2\sqrt{\eta}}{\kappa}
\left[
\sqrt{\frac{\gamma}{\eta}}\bar{\cal H}+\frac{m^2\sqrt{\eta}}{\kappa}
\frac{\gamma}{\eta}
\left(
\frac{1}{2}\eta_{ij}\gamma^{ij}-1
\right)
\right]}\right.\right.\nonumber\\
&-&\left.\left.\frac{m^2\sqrt{\eta}}{\kappa}
\right)
+N^i\bar{\cal H}_i
\right].\label{eq:Ham2}
\end{eqnarray}

We arrived at the Hamiltonian depending on canonical variables 
$\gamma_{ij},\pi^{ij}$, and on the prescribed  metric $\eta_{ij}$ (determined from the fixed spacetime metric and the functions parametrizing hypersurfaces).
The constant provides vacuum energy normalizing, i.e. when the Riemannian metric coincides with the flat one $g_{\mu\nu}=h_{\mu\nu}$, we have ${\rm H}=0$ for any spacelike hypersurfaces.

Let us add that it is possible to do not exclude all constraints from the Hamiltonian and exploit the Dirac brackets from the very beginning. We also can at first make secondary constraints compatible with dynamics of equations
(\ref{eq:he1}) -- (\ref{eq:he2}) -- (\ref{eq:he3}), this will give us a chance to determine the Lagrangian multipliers through canonical variables, but this will not lead us to new constraints.
After substituting these Lagrangian multipliers into equations
(\ref{eq:he2}) -- (\ref{eq:he3}) we will obtain relations which are equivalent to the famous harmonicity conditions
\begin{equation}
D_{\mu}f^{\mu\nu}=0.
\end{equation}
But the Dirac brackets exclude the independent role of these equations and these equations become a consequence of the Hamiltonian equations of motion generated by Hamiltonian  (\ref{eq:Ham2}) and brackets (\ref{eq:DB}).

\section{Scalar field as an example of gravity source}

The formalism developed above will be obviously incomplete without a demonstration of extending it onto matter fields.  Let us suppose that their interaction with gravity is minimal, then the matter fields Lagrangian density  ${\cal L}_M$ will depend on the set of fields $\phi^A(X^{\alpha})$, on their spacetime coordinate derivatives and on metric $g_{\mu\nu}(X^{\alpha})$, transforming as the scalar density under general coordinate transformations. The Hamiltonian form of matter fields action can be obtained through Kucha\u{r}'s procedure~\cite{Kuchar}, and as a result we get
\begin{equation}
S_M=\int\limits^{t_2}_{t_1}dt\int\limits_{R^3}d^3x\left(\pi_A\phi^A_{,0}-
\bar N\bar{\cal H}_M-\bar N^i\bar{\cal H}_{Mi}\right).\label{eq:maction}
\end{equation}
The combination of this action with the gravitational action (\ref{eq:action})
results simply in adding of quantities $\bar{\cal H}_M$ and $\bar{\cal H}_{Mi}$ to
previously obtained $\bar{\cal H}$ and $\bar{\cal H}_i$, which depend on gravitational variables only.

To illustrate this we consider the scalar field with the Lagrangian density
\begin{equation}
{\cal L}_M=-\sqrt{-g}\left(
\frac{1}{2}g^{\mu\nu}\partial_{\mu}\phi\partial_{\nu}\phi+U(\phi)
\right)
\end{equation}
The transformation to 3+1-notations and the Legandre transform may be done separately of the gravitational contribution
\begin{equation}
{\cal L}_M=-N\sqrt{\eta}\left(f^{\perp\perp}\phi_{,\perp}\phi_{,\perp}
+2f^{\perp i}\phi_{,\perp}\phi_{,i}+\frac{1}{f^{\perp\perp}}\left(
f^{\perp i}f^{\perp j}-\frac{\gamma\gamma^{ij}}{\eta}
\right)\phi_{,i}\phi_{,j}+U(\phi)
\right),
\end{equation}
where
\begin{equation}
\phi_{,0}=-N\phi_{,\perp}+N^i\phi_{,i}.
\end{equation}
The momentum is determined as usual
\begin{equation}
\pi_{\phi}=\frac{\partial {\cal L}}{\partial\phi_{,0}}
=\sqrt{\eta}\left(f^{\perp\perp}\phi_{,\perp}-f^{\perp i}\phi_{,i}
\right),
\end{equation}
and the velocity may be expressed through the momentum as follows
\begin{equation}
\phi_{,0}=-\frac{N}{f^{\perp\perp}\sqrt{\eta}}\pi_{\phi}+\left(N^i-
N\frac{f^{\perp i}}{f^{\perp\perp}}\right)\phi_{,i}.
\end{equation}
After providing the corresponding Legandre transform  the action for scalar field will take the form (\ref{eq:maction}), where
\begin{equation}
\bar{\cal H}_M=\frac{1}{\sqrt{\gamma}}\left(
\frac{\pi^2_{\phi}}{2}+\frac{1}{2}\gamma\gamma^{ij}\partial_i\phi\partial_j\phi
+\gamma U(\phi)
\right),\quad
\bar{\cal H}_{Mi}=\pi_{\phi}\phi_{,i}.
\end{equation}
Therefore, after combining actions of the gravitational and the scalar fields we get the same primary constraints
(\ref{eq:constraint1}), (\ref{eq:constraint2}), as in the case of pure gravity, and the full Hamiltonian with constraints  will have the same form (\ref{eq:Ham1}). The procedure of constraints exclusion also does not change and the final form of the Hamiltonian remains the same
(\ref{eq:Ham2}), where now
\begin{eqnarray}
\bar{\cal H}_i&=&-2\pi_{i|j}^j+\pi_{\phi}\phi_{,i},\nonumber\\
\bar{\cal H}&=&\frac{1}{\sqrt{\gamma}}\left(-
\frac{1}{\kappa}\gamma\tilde R+\kappa(\mathrm{Sp}\pi^2
-\frac{\pi^2}{2}) +
\frac{\pi^2_{\phi}}{2}+\frac{1}{2}\gamma\gamma^{ij}\partial_i\phi\partial_j\phi
+\gamma U(\phi) \right),\nonumber\\
\{F,G\}_D&=&\int\limits_{R^3}d^3x\left[\frac{\delta
F}{\delta\gamma_{ij}} \frac{\delta G}{\delta\pi^{ij}}+\frac{\delta
F}{\delta\phi} \frac{\delta G}{\delta\pi_\phi} -\frac{\delta
F}{\delta\pi^{ij}}\frac{\delta G}{\delta\gamma_{ij}} -\frac{\delta
F}{\delta\pi_\phi}\frac{\delta
G}{\delta\phi}\right].\label{eq:DB2}
\end{eqnarray}
It is easy to be convinced that the hamiltonian equations of motion for the system of interacting scalar and gravitational fields are of the following form (in the case  $U(\phi)=1/2M^2\phi^2$):
\begin{eqnarray}
\phi_{,0}&=&\int\limits_{R^3}d^3x\Bigl(
\bar N\{\phi,\bar{\cal H}\}_D+\bar N^k\{\phi,\bar{\cal H}_k\}_D
\Bigr)\nonumber\\
&=&\bar N\frac{\pi_\phi}{\sqrt{\gamma}}+{\bar N}^i\phi_{,i},\\
\pi_{\phi,0}&=&\int\limits_{R^3}d^3x\Bigl(
\bar N\{\pi_{\phi},\bar{\cal H}\}_D+\bar N^k\{\pi_{\phi},\bar{\cal H}_k\}_D
\Bigr)\nonumber\\
&=&(\bar N\sqrt{\gamma}\gamma^{ij}\partial_j\phi)_{,i}-\bar N\sqrt{\gamma}M^2
\phi+(\bar N^i\pi_\phi)_{,i},\\
\gamma_{ij,0}&=&\int\limits_{R^3}d^3x\Bigl(
\bar N\{\gamma_{ij},\bar{\cal H}\}_D+\bar N^k\{\gamma_{ij},\bar{\cal H}_k\}_D
\Bigr)\nonumber\\
&=&{\bar N}_{i|j}+{\bar N}_{j|i}+\kappa\frac{2\bar N}{\sqrt{\gamma}}(\pi_{ij}
-\gamma_{ij}\frac{\pi}{2}),\\
\pi^{ij}_{,0}&=&\int\limits_{R^3}d^3x\Bigl(
\{\pi^{ij},\bar N\bar{\cal H}\}_D+\bar N^k\{\pi^{ij},\bar{\cal H}_k\}_D
\Bigr)\nonumber\\
&+&\frac{m^2}{\kappa}\bar N\sqrt{\gamma}\left[
\gamma^{ij}+\frac{1}{2}\eta_{kl}\left(
\gamma^{ki}\gamma^{lj}-\gamma^{ij}\gamma^{kl}
\right)
\right]\nonumber\\
&=& -\frac{1}{2}\bar N\sqrt{\gamma}(\gamma^{ij}\gamma^{mn}-\gamma^{im}
\gamma^{jn})\partial_m\phi\partial_n\phi-\frac{1}{2}\bar N\sqrt{\gamma}
\gamma^{ij}M^2\phi^2\nonumber\\
&-&\frac{1}{\kappa}\bar N\sqrt{\gamma}(R^{ij}-\gamma^{ij}R)+\kappa\frac{\bar
N}{\sqrt{\gamma}}(\pi\pi^{ij}-2\pi^{ik}\pi^j_k)\nonumber\\
&+&\frac{1}{\kappa}
\sqrt{\gamma}({\bar N}^{|ij}-\gamma^{ij}\bar N^{|k}_{|k})+
(\pi^{ij}{\bar N}^k)_{|k}-\pi^{ik}{\bar N}^j_{|k}-\pi^{kj}{\bar
N}^i_{|k}\nonumber\\
&+&\frac{m^2}{\kappa}\bar N\sqrt{\gamma}\left[
\gamma^{ij}+\frac{1}{2}\eta_{kl}\left(
\gamma^{ki}\gamma^{lj}-\gamma^{ij}\gamma^{kl} \right) \right].
\end{eqnarray}
It is evident that the difference with the corresponding equations of General Relativity appears in the last equation only and has the order of magnitude $O(m^2/\kappa)$. The dependence of quantities $f^{\perp\perp}$, $f^{\perp
i}$ on variables (\ref{eq:f_perp}), whose Hamiltonian equations have been derived above, may be ignored in the calculating of Dirac brackets acording to formulas
(\ref{eq:secondconstraints}).

\section{Poincar\'e group in Hamiltonian formalism of the RTG}
Among all variants of the hamiltonian evolution arising from arbitrariness of functions $N(x)$, $N^i(x)$ Hamiltonian~(\ref{eq:Ham2}) contains special transformations preserving the Minkowski metric. So, if we take hyperplanes as hypersurfaces  and if we take Cartesian systems of coordinates on them, we will have on these hypersurfaces metric $\eta_{ij}$ induced by Minkowski metric of spacetime (7) in the simplest form $\eta_{ij}=\delta_{ij}$, and we will have transformation functions~(\ref{eq:parameters}) as follows
\begin{equation}
N=A_kx^k+a,\quad N^i=A_{ik}x^k+a^i,\label{eq:Poincare}
\end{equation}
where
\[A_{ik}=-A_{ki}.\]
Then Hamiltonian~(\ref{eq:Ham2}), according to its linearity in functions $N(x)$, $N^i(x)$, will take a form
\begin{equation}
H=P^0a-P^ia^i+M^kA_k+\frac{1}{2}M^{ik}A_{ik},
\end{equation}
where
\begin{eqnarray}
P^0 &=&-\frac{m^2}{\kappa}\int \left(1+f^{\perp\perp}\right)d^3x,\nonumber \\
P_i &=& -\frac{m^2}{\kappa}\int f^{\perp i}d^3x\equiv-\int{\cal H}_i d^3x,\nonumber \\
M^{ik} &=& -\frac{m^2}{\kappa} \int\left(x^i f^{\perp k}-x^k f^{\perp i}\right)d^3x\equiv\int\left(x^k{\cal H}_i-x^i{\cal H}_k\right)d^3x,\nonumber \\
M^k &=& -\frac{m^2}{\kappa}\int x^k(1+f^{\perp\perp})d^3x.
\end{eqnarray}
The meaning of these operators is clear as they are special cases of the general Hamiltonian and correspond to different choices  of coordinate transformations which they generate: $P^0$ refers to the time translation, $P^i$ -- to the spatial translations, $M^{ik}$ -- to the spatial rotations and $M^k$ -- to Lorentz boosts.
Our notations are chosen in order to make easy comparison with the analogous formulas from paper~\cite{RT}, where Poincar\'e algebra has been considered in the asymptotically flat spacetime of General Relativity.

But before reducing our Hamiltonian to such restricted form it is useful to derive the algebra of Dirac brackets (\ref{eq:DB2}) for the general Hamiltonians. Let us take
\begin{eqnarray}
{\rm H}&=&\int\limits_{R^3}d^3x\Biggl(N{\cal H}+N^i{\cal
H}_i\nonumber
\\&=&\int\limits_{R^3}d^3x\Biggl(
\bar N\bar {\cal H}+\bar N^i\bar{\cal H}_i\nonumber\\
&+&\frac{m^2\sqrt{\eta}}{\kappa}N\left[
-1-\frac{f^{\perp\perp}}{2}+ \frac{f^{\perp i}f^{\perp
j}\eta_{ij}}{2f^{\perp\perp}}-
\frac{1}{f^{\perp\perp}}\frac{\gamma}{\eta}\left(\frac{1}{2}
\eta_{ij}\gamma^{ij}-1 \right) \right]\Biggr),\label{eq:Ham3}
\end{eqnarray}
where we treat $f^{\perp\perp}$, $f^{\perp i}$ as functions having zero Dirac brackets. This is justified by the well-known statement that second class constraints may be taken into account both before and after calculations of Dirac brackets.
So, we will account for the first half of constraints (\ref{eq:constraint1}),
(\ref{eq:constraint2}) before, and the second half 
(\ref{eq:f_i}), (\ref{eq:f_perp}) -- after. As usual all surface integrals in integration by parts are discarded. It is reasonable for island systems when radiation is allowed only inside of them, not at infinity where the pseudo-Riemannian metric tends to the flat one with the rate given by the Yukawa behaviour.

The results of calculations may be presented in the form suitable for comparison with the analogous formulas of GR:
\begin{eqnarray}
\{H(\alpha, \alpha^i),H(\beta,\beta^j)\}&=&\int d^3x \Bigl[\bar\lambda\bar{\cal H}+\bar\lambda^k\bar{\cal H}_k+(\bar\alpha\bar\beta^k_{|k}-\bar\beta\bar\alpha^k_{|k})\bar{\cal H}\nonumber\\&-&\frac{m^2}{\kappa}\sqrt{\gamma}\gamma_{k\ell}(\bar\alpha\bar\beta^{k|\ell}-\bar\beta\bar\alpha^{k|\ell})
(2-\eta_{mn}\gamma^{mn}))\nonumber\\
&-&\frac{m^2}{\kappa}\sqrt{\gamma}\eta_{k\ell}(\bar\alpha\bar\beta^{k|\ell}-\bar\beta\bar\alpha^{k|\ell})\label{eq:algebra_g}
\Bigr],\nonumber\\
\bar\lambda&=&\bar\alpha^i\bar\beta_{,i}-\bar\beta^i\bar\alpha_{,i},\nonumber\\
\bar\lambda^k&=&\gamma^{k\ell}(\bar\alpha\bar\beta_{,\ell}-\bar\beta\bar\alpha_{,\ell})+\bar\alpha^\ell\bar\beta^k_{,\ell}-\bar\beta^\ell\bar\alpha^k_{,\ell},\label{eq:algebra_par}
\end{eqnarray}
or in the form corresponding to parametrized field theories at the background of flat metric:
\begin{eqnarray}
\{H(\alpha, \alpha^i),H(\beta,\beta^j)\}&=& H(\lambda,
\lambda^k)+\int \frac{\partial {\cal H}}{\partial \eta_{ij}}
\left(\alpha{\cal L}_{\vec\beta}\eta_{ij}-\beta{\cal
L}_{\vec\alpha}\eta_{ij}\right)d^3x,\label{eq:algebra_h}\nonumber\\
\lambda&=&\alpha^i\beta_{,i}-\beta^i\alpha_{,i},\nonumber\\
\lambda^k&=&\eta^{k\ell}(\alpha\beta_{,\ell}-\beta\alpha_{,\ell})+\alpha^\ell\beta^k_{,\ell}-\beta^\ell\alpha^k_{,\ell},
\end{eqnarray}
where ${\cal L}_{\vec\alpha}\eta_{ij}$ is the Lee derivative of metric
$\eta_{ij}$ in the direction of vector field  $\vec\alpha$.
The difference from GR appears in Eq.(\ref{eq:algebra_g}) both in terms proportional to the graviton mass, and in the coefficient standing before $\bar{\cal H}$. The last fact is resulting from the proportionality of $\bar N$ to 
$\sqrt{\gamma}$. Equations (\ref{eq:algebra_g}), so, do not reproduce the famous hypersurface deformation algebra~\cite{Dirac},\cite{RT}, because functions $\bar N$,
$\bar N^i$ do not coincide with parameters of this algebra.

The substitution into Eqs.~(\ref{eq:algebra_h}) expressions of the form (\ref{eq:Poincare}), corresponding to Poincar\'e transformations, in the place of arbitrary functions $\alpha$, $\alpha^i$, $\beta$, $\beta^j$, results in Poincare algebra relations for the Dirac brackets:
\begin{eqnarray}
  \{P^0,P_i\}_D &=& 0, \quad  \{P_i,P_j\}_D = 0,\nonumber \\
  \{P^0,M^{ik}\}_D &=& 0, \quad  \{P_i,M^{jk}\}_D = \delta_{ik}P_j-\delta_{ij}P_k,\nonumber \\
  \{M^{ij},M^{k\ell}\}_D
&=&\delta_{ik}M^{j\ell}-\delta_{i\ell}M^{jk}+\delta_{j\ell}M^{ik}-\delta_{jk}M^{i\ell},\nonumber \\
  \{P^0,M^i\}_D &=& -P^i, \quad  \{P_i,M^j\}_D =  -\delta_{ij}(P^0-c^0),\nonumber\\
 \{M^k,M^{ij}\}_D&=&\delta_{kj}(M^i-c^i)-\delta_{ki}(M^j-c^j),\quad \{M^i,M^j\}_D=-M^{ij}.
\end{eqnarray}
Additive terms $c^0=m^2/\kappa\int d^3x$ and$c^i=m^2/\kappa\int x^id^3x$ in $P_0$ and $M^i$, which do not depend on canonical variables and which are given by divergent spatial integrals play the role of central charges in Poincar\'e algebra and correspond to the classical renormalization of the vacuum energy.
They arise due to our intention to provide strictly zero energy for the empty Minkowskian spacetime. For this purpose a zero order over physical field term is introduced into the Hamiltonian (and also into the Lagrangian). As to linear over physical field terms in the Lagrangian and Hamiltonian, they appear only as total derivatives, and so do not contribute to the equations of motion.

In principle, there is another opportunity: we can discard constant terms from the definitions of Poincar\'e generators:
\begin{eqnarray}
P^0 &=&-\frac{m^2}{\kappa}\int f^{\perp\perp}d^3x, \nonumber\\
P_i &=& -\frac{m^2}{\kappa}\int f^{\perp i}d^3x\equiv-\int{\cal H}_i d^3x,\nonumber \\
M^{ik} &=& -\frac{m^2}{\kappa} \int\left(x^i f^{\perp k}-x^k f^{\perp i}\right)d^3x\equiv\int\left(x^k{\cal H}_i-x^i{\cal H}_k\right)d^3x, \nonumber\\
M^k &=& -\frac{m^2}{\kappa}\int x^k f^{\perp\perp}d^3x.\label{eq:nocentral}
\end{eqnarray}
Then we get Poincar\'e algebra without central charges:
\begin{eqnarray}
  \{P^0,P_i\}_D &=& 0, \quad  \{P_i,P_j\}_D = 0,\nonumber \\
  \{P^0,M^{ik}\}_D &=& 0, \quad  \{P_i,M^{jk}\}_D = \delta_{ik}P_j-\delta_{ij}P_k,\nonumber \\
  \{M^{ij},M^{k\ell}\}_D
&=&\delta_{ik}M^{j\ell}-\delta_{i\ell}M^{jk}+\delta_{j\ell}M^{ik}-\delta_{jk}M^{i\ell},\nonumber \\
  \{P^0,M^i\}_D &=& -P^i, \quad  \{P_i,M^j\}_D =  -\delta_{ij}P^0,\nonumber\\
 \{M^k,M^{ij}\}_D&=&\delta_{kj}M^i-\delta_{ki}M^j,\quad \{M^i,M^j\}_D=-M^{ij}.
\end{eqnarray}
But in this case the energy density for Minkowski space without matter will have a nonzero and positive value $\frac{m^2}{\kappa}$,
and the total energy will be infinite:
\begin{equation}
P^0 =-\frac{m^2}{\kappa}\int f^{\perp\perp}d^3x=\frac{m^2}{\kappa}\int d^3x.
\end{equation}

We can conclude that it is natural to consider tensor
\begin{equation}
T^{\mu\nu}_{total}=-\frac{m^2}{\kappa}f^{\mu\nu}\equiv-\frac{m^2}{\kappa}\frac{\sqrt{-g}}{\sqrt{-h}} g^{\mu\nu},\label{eq:EL}
\end{equation}
as a total energy-momentum tensor of gravitational field and matter. Let us add some reasons for this.
The variational principle for RTG Lagrangian leads us to the following equations:
\begin{equation}
R^{\mu\nu}-\frac{1}{2}g^{\mu\nu}R+\frac{m^2}{2}\left[h_{\alpha\beta}\left(\frac{1}{2}g^{\mu\nu}g^{\alpha\beta}-g^{\alpha\mu}g^{\beta\nu} \right)-g^{\mu\nu}  \right]=\frac{\kappa}{2}T^{\mu\nu}_{matter},
\end{equation}
where $T^{\mu\nu}_{matter}$ is the  energy-momentum tensor of matter
\begin{equation}
T^{\mu\nu}_{matter}=\frac{2}{\sqrt{-g}}\frac{\delta S_M}{\delta g_{\mu\nu}}.
\end{equation}
If we take a covariant derivative $\nabla_\nu$, which is compatible with metric $g_{\mu\nu}$, of both sides of Eq.(\ref{eq:EL}) we obtain (compare with Ref.~\cite{Log} Eq.(5.17))
\begin{equation}
\left(T^{\mu\nu}_{total}\right)_{;\nu}=\frac{\sqrt{-g}}{\sqrt{-h}}\nabla_\nu T^{\mu\nu}_{matter},
\end{equation}
where the semicolon denotes the covariant derivative compatible with the flat background metric. The matter equations of motion leads to equations
\begin{equation}
\nabla_\nu T^{\mu\nu}=0,
\end{equation}
so, if the gravitational field equations (\ref{eq:EL}) are also valid, then we have
\begin{equation}
\left(T^{\mu\nu}_{total}\right)_{;\nu}=0.
\end{equation}
Then every Killing vector $\xi^{\mu(i)}$, ($i=1$,2,\ldots,10) of the Minkowski metric
\begin{equation}
\xi^{(i)}_{\mu;\nu}+\xi^{(i)}_{\nu;\mu}=0,
\end{equation}
provides us with a conserved quantity
\begin{equation}
I[\Sigma,\xi^{(i)}]=\int\limits_\Sigma T^{\mu\nu}_{total}\xi_{\mu}^{(i)}dS_\nu,
\end{equation}
in correspondence with Eqs.(\ref{eq:nocentral}). The energy density (Eq.(57) or Eq.(62)) sign problem requires further study. Treated in the linear approximation the energy density is an indefinite quadratic form, up to the spatial divergence. But only the full theory has an essential meaning. If some pathology in the form of negative energy flow generated by scalar component were present in the theory, then it should appear, for example, in the form of spherial waves radiation. But in publication~\cite{LM} (see also Chapter 12 of Ref.~\cite{Log}) it was shown that the radiation of the gravitational field scalar component is abcent in spherically symmetric case and the external field is only static. Earlier analogous results were derived in Ref.~\cite{Losk}

\section{Conclusion}

Let us summarize our results, formulate some expectations for the future and make some proposals. Up to now the massive gravity has not been considered in spacetime covariant Hamiltonian formalism (with arbitrary lapse and shift). As a result, Hamiltonians treated before did not provide freedom for the spacetime foliations. This work ensures us the Hamiltonian formalism flexibility which can be used to clear such fundamental issues as the causality in bimetric theory (see also bigravity~\cite{DR-D}), as stability of the solutions, as positivity of energy and so on. In paricular, it is often required to compare predictions of massless and massive gravity for the same physical problem. As canonical variables and their brackets coincide, the main difference consists in four constraints which present in massless case and absent in massive one. Therefore, it would be rather interesting to compare the Cauсhy problems with the same initial conditions for the two theories on rather long time interval, for example, compared with the Solar system age. We hope that the above mentioned problems are possible to study  with modern numerical methods of computer calculations.


\end{document}